\documentclass[12pt]{article}
\usepackage{pic02}
\usepackage{hyperref}
\usepackage{url}
\usepackage{graphicx}

\begin{document}

\title{\bf {BABAR LEVEL 1 DRIFT CHAMBER TRIGGER UPGRADE}
\author{Valerie Halyo \\
{\em Stanford Linear Accelerator Center,
	 2575 Sand Hill RD, Menlo Park CA, USA}}}
\maketitle

\baselineskip=14.5pt
\begin{abstract}
As PEP-II is exceeding the original design luminosity, 
BaBar is currently upgrading its Level 1 Drift Chamber
Trigger (DCT) to reduce the rate of background Level 1 triggers 
by more than $50\%$ while preserving the high Level 1 trigger physics efficiency.
New Z-Pt-Discriminator VME boards (ZPD) utilizing the stereo hit
information from the drift chamber are being built to extract the track
z coordinate at the beam line with a resolution of a few centimeters.
\end{abstract}

\baselineskip=17pt

\section{\bf{Introduction}}
As PEP-II exceeds its original design luminosity, 
BaBar upgrades its Level 1 Drift Chamber Trigger (DCT) to reduce the rate 
of background Level 1 triggers.
The Level 1 trigger system in BaBar receives information from the
Drift Chamber, the Calorimeter, and Instrumented Flux Return. 
The Drift Chamber information sent to the DCT is used to generate segments in
the Track Segment Finder (TSF). This technique uses the drift time in 
order to establish accurate position and time of the track segment.
In addition a Look Up Table derived from data is used to calibrate
the best segment pattern into its corresponding track position and time.
Once the segments are generated they are sent to the 
Binary Link Track Module (BLT) and to the new Z-PT Discriminator (ZPD).
The BLT is responsible to combine these segments into tracks providing the first 
track multiplicity count.
The ZPD extracts the track Pt and z coordinate at the beam line.
The processed data is sent to the GLobal Trigger. 
The  GLT attempts to match the spatial and angular location of calorimeter towers
 and drift chamber tracks. 
This result in $24$  trigger lines which are sent to the Fast control
to issue a L1 trigger.

\section{\bf{The Z-PT Discriminator upgrade}}
The New Z-Pt-Discriminator boards (ZPD) replace the Pt Discriminator boards
which only extracted the Pt of the tracks.
The ZPD boards utilize the stereo hit
information from the drift chamber to extract in addition the track
z coordinate at the beam line.
The ZPD boards select tracks based on their transverse momentum and their origin along the beam
line, $z_0$, in order to reject beam pipe interactions far from the interaction point.
The ZPD algorithm is composed of two main modules the Seed Track Finder and a $z_0$ Fitter.
The first is a pattern recognition module which finds seed tracks with up
to ten segments each, and the second module fits the seed tracks
to a helix in order to determine their $z_0$.
The resulting new trigger line can be based on any desired combination of $z_0$,
 PT and $\tan \lambda$ values.
There are $8$ new VME boards, each one receives input at 1~Gbyte/sec.
The 153-line internal signal mega-bus transmit data at 120~Mhz.  
The are six algorithm engines (XC2V4000 FPGAs)  per ZPD board working in parallel.

\section{\bf{ZPD performance}}

\begin{figure}[htbp]
\begin{center}
\includegraphics[height=6cm,width=7cm]{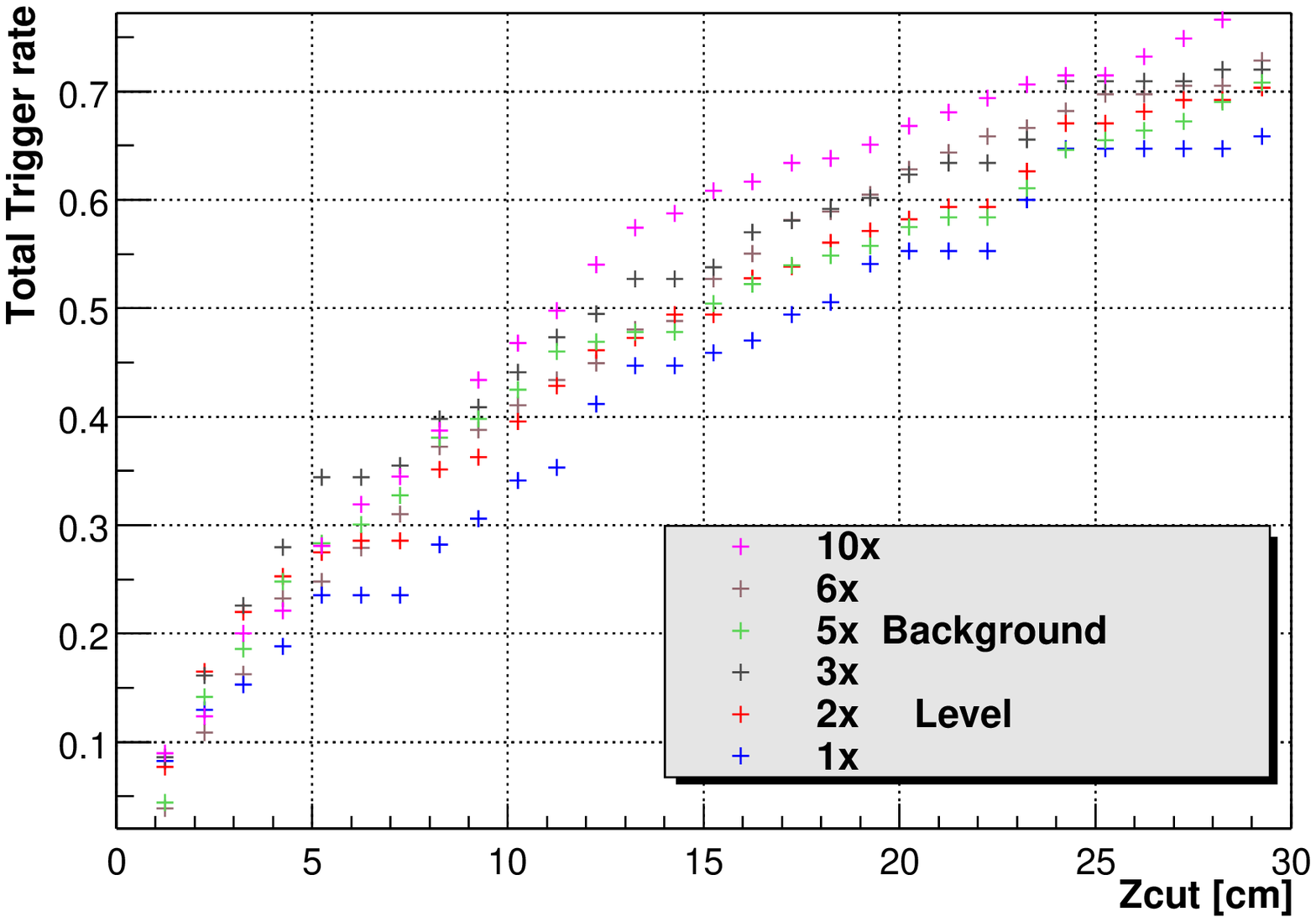}
\hspace{0.25in}
\includegraphics[height=6.2cm,width=7cm]{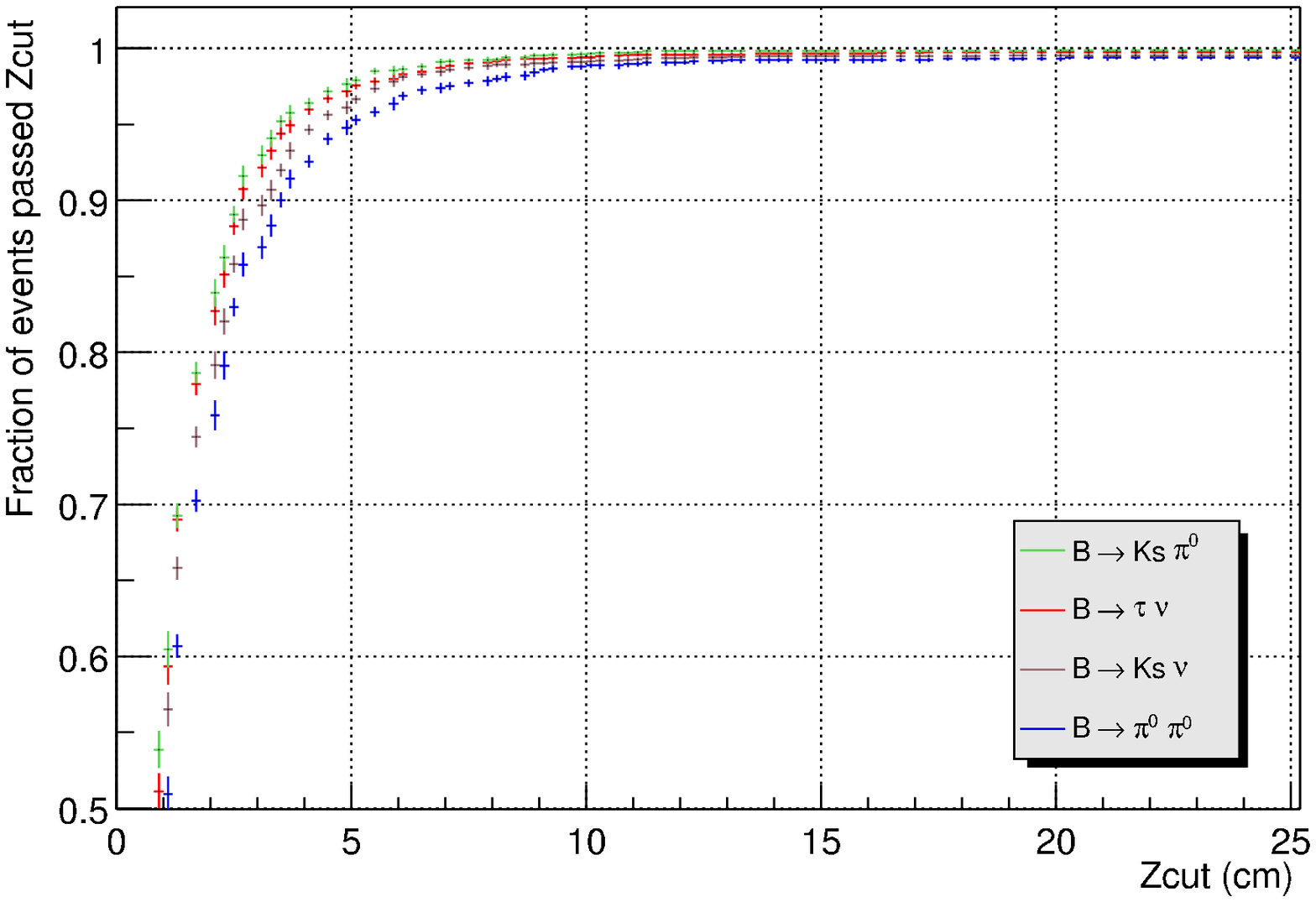}
\caption{The total trigger rate vs. the $|z_0|$ cut in different background 
conditions (Left). Event efficiency for rare $B$ decay samples, where the event 
efficiency is determined by whether the event has at least one 
track with $|z_0|< \mathrm{Zcut}$ (Right).}
\label{fig:efficiency}
\end{center}
\end{figure}

The new ZPD boards are designed to accommodate 
the upcoming BaBar demands for high luminosity.
The ZPD reduces the rate of background Level 1 triggers by more
than $50\%$ while preserving the high Level 1 trigger physics efficiency.
A Monte Carlo simulation with higher background level was generated to test the performance
of the ZPD. The results, as can be seen in fig~\ref{fig:efficiency} demonstrate
that even up to $6$ times the current background level we can achieve about $40\%$
background suppression by rejecting events that lack at least one track 
with $|z_0|<13 \mathrm{cm}$.
In order to test the MC event efficiency using the ZPD algorithm.
The event efficiency was tested on different MC samples including rare B decays.
The denominator is simply all MC events passing at least one existing L1 trigger. 
Fig.~\ref{fig:efficiency} shows the event efficiency of the following 
rare $B$ decays samples, where one $B$ decayed generically
while the other $B$ decayed to:     
$B^0\rightarrow K^0_s\pi^0$, $B^-\rightarrow \tau\nu$,    
$B^0\rightarrow K^0_s\nu\overline{\nu}$, $B^0\rightarrow\pi^0\pi^0$.
The results are very encouraging given the current 
pure DCT triggers are ranging from 80--95\% for these events. 

The total trigger rate reduction due to an addition of a  Z requirement 
$|z_0|<15 \mathrm{cm}$ is $30\%$, and improves to $38\%$ for  $|z_0|<10 \mathrm{cm}$.
However, it should be noted that 
the total trigger rate when the DCT lines are disabled is about $40\%$
which is an irreducible rate (in the current BaBar trigger configuration) since 
they are triggered by at least one EMT-only trigger lines, independent of DCT.

\section{Acknowledgments}
This work was supported by Department of Energy contract DE-AC03-76SF00515.

\end{document}